\documentclass[fleqn,10pt,twocolumn]{article}
\usepackage{graphicx}
\usepackage{amsmath,amssymb}
\usepackage{mathtools}
\usepackage{arydshln}
\usepackage{authblk}
\usepackage{hyperref}
\usepackage[left=2cm,%
                right=2cm,%
                top=2.25cm,%
                bottom=2.25cm,%
                headheight=12pt,%
                letterpaper]{geometry}%
\usepackage[font=small,labelfont=bf,
   justification=justified,
   format=plain, figurename=Figure]{caption}

\usepackage{natbib}
\setcitestyle{authoryear, round, aysep={,}}   

\title{Evolution of family systems and resultant socio-economic structures}
\date{}
\author[1]{Kenji Itao}
\author[1, 2, *]{Kunihiko Kaneko} 

\affil[1]{Department of Basic Science, Graduate School of Arts and Sciences, University of Tokyo, Komaba 3-8-1, Meguro-Ku, Tokyo 153-8902, Japan.}
\affil[2]{Research Center for Complex Systems Biology, University of Tokyo, Komaba 3-8-1, Meguro-Ku, Tokyo 153-8902, Japan.}

\affil[*]{kaneko@complex.c.u-tokyo.ac.jp}

\begin{document}

\flushbottom
\maketitle

\begin{abstract}
Families form the basis of society, and anthropologists have observed and characterised a wide range of family systems. This study developed a multi-level evolutionary model of pre-industrial agricultural societies to simulate the evolution of family systems and determine how each of them adapts to environmental conditions and forms a characteristic socio-economic structure. In the model, competing societies evolve, which themselves comprise multiple evolving families that grow through family labour. Each family has two strategy parameters: the time children leave the parental home and the distribution of inheritance among siblings. The evolution of these parameters demonstrates that four basic family systems emerge; families can become either nuclear or extended, and have either an equal or strongly biased inheritance distribution. Nuclear families in which children leave the parental home upon marriage emerge where land resources are sufficient, whereas extended families in which children staying at the parental home emerge where land resources are limited. Equal inheritance emerges where the amount of wealth required for a family to survive is large, whereas strongly biased inheritance emerges where the required wealth is small. Furthermore, the frequency of polygyny is low in the present model of agricultural societies, whereas it increases for the model of labour-extensive subsistence patterns other than agricultural societies. Analyses on the wealth distribution of families demonstrate a higher level of poverty among people in extended families, and that the accumulation of wealth is accelerated in families with strongly biased inheritance. By comparing wealth distributions in the model with historical data, family systems are associated with characteristic economic structures and then, modern social ideologies. Empirical data analyses using the cross-cultural ethnographic database verify the theoretical relationship between the environmental conditions, family systems, and socio-economic structures discussed in the model. The theoretical studies made possible by this simple constructive model, as presented here, will integrate the understandings of family systems in evolutionary anthropology, demography, and socioeconomic histories.
\end{abstract}

\section{Introduction}
Families are basic components of society. Studies of family and kinship lay at the core of anthropology \citep{fox1983kinship, levi1958anthropologie, harrell1997human, shenk2011rebirth}.
Families comprise people who are connected by three basic relationships, i.e. husband--wife (or generally, marital partners'), parent--child, and inter-sibling relationships \citep{white1963anatomy}, and a variety of rules (or patterns) are observed concerning them \citep{harrell1997human,laslett1988family, todd1999diversite}.
Cultural traits pertaining to family relationships are slow to change, because they tend to be inherited vertically from parent to child, and are regulated by social norms \citep{cavalli1981cultural}. 
Indeed, empirical studies have revealed such slow changes based on population history \citep{guglielmino1995cultural, mulder2001study, minocher2019explaining}. Family traits have also attracted significant attention as basic factors of social characteristics in the study of history \citep{braudel1992civilization1, braudel1992civilization2, braudel1992civilization3} and historical demography \citep{macfarlane2002savage}, among others.

The appearance of these traits was previously explained by cultural transmission or adaptation to social and ecological conditions \citep{todd2011origine, laslett2015world, goldschmidt1971structure}.
Social studies unveiled a correlation between the period of exposure to the Western Church and the emergence of nuclear families \citep{schulz2019church}.
In evolutionary anthropology, relationships between family members are explained based on parental investment theory and intra-family competition for reproductive resources \citep{trivers1974parent, ji2014intergenerational}.
Subsistence patterns and other socio-ecological conditions
that affect family traits have recently been revealed quantitatively \citep{colleran2014farming, macfarlan2019emergent, gibson2011land, ross2018greater}. 
Phylogenetic comparative analyses have been performed to infer the origin and historical change of family traits by controlling the statistical non-independence due to the shared ancestry \citep{fortunato2006bridewealth, fortunato2010your, holden2003spread, mulder2001study, minocher2019explaining}.
In particular, they revealed that the presence of heritable resources that are typically observed in agricultural/ pastoralist society leads to sibling competition over inheritance \citep{gibson2011land} and to patriliny \citep{holden2003spread}. Differentiation of very rich elite from the majority in agricultural society leads to a lower frequency of polygyny \citep{ross2018greater}.

Indeed, correlations have been observed between family traits and several socio-ecological conditions.
However, it is unclear whether social factors determine family traits or vice versa \citep{mace2011macro}.
The reverse effect from family traits to social conditions is also reported in historical demography \citep{macfarlane2002savage}. 

As two characteristics in the family system, we consider parent--child and inter-sibling relationships.
Although a variety of characteristics can represent these relationships, we focus on residence and inheritance patterns. The residence pattern can refer to nuclear families that comprise a pair of parents and their unmarried children, or to extended families that can involve parents and their married children. The inheritance pattern can refer to either the equal or strongly biased distribution of inheritance among siblings.
On this basis, we can suppose four ideal types.
(1) Absolute nuclear families, which are nuclear families with unequal inheritance. (2) Egalitarian nuclear families, which are nuclear families with equal inheritance. (3) Stem families, which are extended families with unequal inheritance. (4) Community families, which are extended families with equal inheritance.
Indeed, other characteristics of family systems, such as marriage patterns, would be necessary to classify family systems comprehensively, which needs future works. Nevertheless, there is notable variation in the residence and inheritance patterns in pre-industrial agricultural societies in Europe, Northern Africa, and Asia, among others \citep{harrell1997human}.
The influence of agricultural societies on political systems in the modernising era has been described \citep{wallerstein2011modern, rosener1993bauern}.
The link between the above four family systems and modern social ideologies has been discussed \citep{laslett1988family}; Liberalism, liberal egalitarianism, social democracy, and communism are dominant in regions with absolute nuclear, egalitarian nuclear, stem, and community families, respectively \citep{todd1990invention, todd1999diversite}.
However, the discussion regarding the relationship between family systems and ideologies remains largely psychoanalytical. Hence, theoretical studies to unveil the conditions of the evolution of each family system and to connect family systems with socio-economic structures need to be conducted.

To consider the interaction of ecological conditions, family systems, and social structures, a constructive approach to reveal the relationships between them is required.
In addition, given that family traits are inherited from parents with slight changes over generations, it is appropriate to model their long-term evolution through the accumulation of small variations, as represented by mutations.
By modelling the evolution of family systems to adapt to socio-ecological conditions, which in turn form the society-level economic structure, we aim to integrate the understandings in evolutionary anthropology, demography, and socioeconomic histories.
To model the economic consequence of family behaviour, we focus on agricultural society, where family systems determine residence and inheritance patterns in land usage \citep{todd1990invention, todd2011origine}.
Children may cultivate lands of their own or work together on their parents' land. One heir may inherit the land and most of the property exclusively, or the land and property may be divided equally among family members \citep{berkner1976inheritance, kaser2002power}. Characteristics of the pre-industrial agricultural society include the importance of human labour, land and property in production \citep{colleran2014farming}, a positive correlation of wealth and the number of offspring \citep{gibson2011land}, and the diminishing returns to labour input \citep{ricardo1891principles, evenson2001effect}. Hereby, we build the minimal model that is appropriate as long as these conditions are satisfied.

Notably, families constitute society, whereas society, as well as ecology, provides the environment for families.
Hence, our model adopts a framework involving multi-level evolution for a hierarchical system. 
The multi-level evolution was originally introduced to explain the evolution of cooperative behaviour among eusocial insects by examining the conflict between the fitness of an individual and that of a group \citep{wilson1997altruism, wilson2007rethinking}. This framework is generally applied to the evolution of group-level structure in hierarchical systems \citep{traulsen2006evolution, takeuchi2017origin, spencer2001multilevel, turchin2009evolution}.
In the previous study, the framework was applied to construct a mathematical model for the evolution of kinship structures in clan societies, which revealed the environmental dependencies of diverse kinship structures \citep{itao2020evolution, itao2021emergence}. The variety of family traits regarding cousin marriage preferences and clan exogamy, as well as descent systems, were investigated in detail therein. Here, in contrast, we mainly focus on parent--child residence patterns and inter-sibling inheritance patterns, and briefly mention the conditions for polygyny.

In this study, we investigate the evolution of family systems and social structures by introducing an agent-based multi-level evolutionary model of pre-industrial agricultural societies. Competition is considered at two levels: that of family, which is an individual agent of the model, and society, which is a group of families.
Families produce wealth through family labour and reproduce their population.
They possess two strategy parameters concerning the time children leave their parents' home and the distribution of inheritance among them. These parameters are transmitted with slight mutations in each generation.
Evolutionary simulations show that four family systems emerge depending on environmental parameters that characterise the land scarcity and perturbations that damage society.
Then, the model is extended by adding the marriage process. We show that this extension affects the above result only minimally, whereas it facilitates the discussion of conditions for son-biased investment and polygyny.
We then describe the characteristics of social structure in terms of the distribution of wealth in society and relate them to family systems.

Finally, the theoretical results are verified through a data analysis using the standard cross-cultural sample (SCCS), a global ethnographic database of premodern societies \citep{murdock1969standard, kirby2016d}. SCCS contains data from 186 societies, which are thought to be culturally and linguistically independent of each other. In the discussion section, we show the relationships between family systems, socioeconomic structures, and the development of political ideology in the modernising era, by referring to socioeconomic histories.

\section{Model}
\subsection{Overview of the model}
\begin{figure}[tb]
    \centering
    \includegraphics[width=\linewidth]{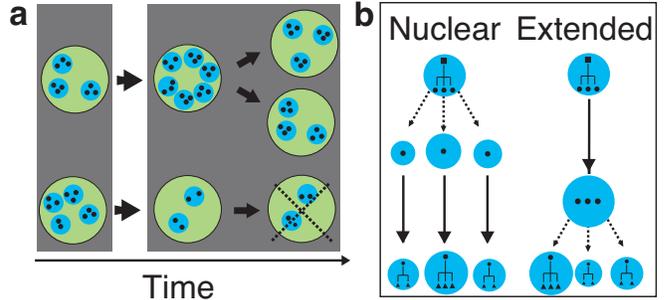}
    \caption{Schematic of the model. \textbf{a,} The multi-level evolutionary process. Societies (green) comprise families (blue) in which family members (black) live together. The grey frame represents a single generation. When the population of a society reaches twice the initial population, a society splits and another society is removed from the system at random to keep the number of societies fixed --- i.e. we adopt the hierarchical Moran process. \textbf{b,} The life cycle of families. The black squares, circles, and triangles represent members of different generations. The size of the blue circle reflects the amount of wealth that each family possesses. The dashed arrows show the separation of siblings, and the arrows show the production of wealth through family labour. Each family has a strategy parameter $s$. With the probability of $1 - s$, siblings are separated before production, and then they produce independently and reproduce the next generation, as in a nuclear family. On the other hand, with the probability of $s$, siblings produce together, after which they build their own home and reproduce the next generation, as in an extended family. When we extend the model to consider marriage, its process is added just before production.
    }
    \label{fig:model_scheme}
\end{figure}

The model is described below in general terms in this subsection (see the following subsection for the mathematical formulation).
A schematic of the model is shown in Figure \ref{fig:model_scheme}a,b. Society consists of families. In each family, individual members live and work together. Children build families of their own by inheriting their parents' wealth. Wealth $w$ is accumulated through production, which in turn increases the level of production and the population, as empirically reported \citep{macfarlane2002savage, gibson2011land}.
Each society splits into two when the number of families therein doubles its initial value $N_f$, and each family is randomly assigned to one of the two daughter societies.
At this time, another society is removed at random so that the number of societies in the entire system remains fixed to $N_s$. This process can be interpreted as invasion, imitation, or the coarse-grained description of a growing system. Therefore, societies that grow faster replace others, resulting in society-level evolution. This multi-level evolution of families and societies follows the hierarchical Moran process \citep{itao2020evolution, traulsen2006evolution, takeuchi2017origin}.

In each simulation, the environmental parameters are given. Of particular importance are the following two:
the capacity $c \times N_f$, which represents the amount of available land resources in society (hereafter called land capacity), and $\epsilon$, which is wealth required for a family to survive the perturbations that damage society (hereafter called wealth required for survival).
Agricultural production depends on the amount of available land resources, whereas insufficient capacity limits land resources per family.
In this model, when the number of families in a society exceeds the capacity, the land resources and the production rate for each family decrease inversely with the number of families at that time.
As for the wealth required for survival, $\epsilon$ must be paid by a family at the moment of its building.
If wealth $w$ is less than $\epsilon$, its members die without reproducing.

Families have population and wealth, as well as the two strategy parameters, i.e. $\lambda$, which represents the inequality in the inheritance of wealth among siblings, and $s$, which represents the probability of children staying at their parents' home to produce together. The $m$th child's share of the inheritance is proportional to $\exp (-\lambda m)$. Therefore, $\lambda = 0$ represents an equal division of inheritance, whereas a larger $\lambda$ represents the eldest child inheriting more. In some societies, the youngest child inherits the most instead of the eldest \citep{todd2011origine}. If necessary, the order of children could be arranged in reverse to include such a case.
Each family forms an extended family with probability $s$ or otherwise forms a nuclear family (explained in detail below).
These parameters may be determined by intra-family competition among parents and siblings\citep{trivers1974parent, ji2014intergenerational}. Here, we do not model the competition explicitly but do so implicitly by tracing the evolution of $\lambda$ and $s$.

\begin{table}[tb]
\caption{Parameters used in the model. 
$\ast$ represents the variables varied in Figure \ref{fig:phase_phase_structure}. $\ast\ast$ represents the variables varied in Figure S1.
} 
  \label{table:param}
 \centering
 \resizebox{\linewidth}{!}{
    \begin{tabular}{l|l|l} 
Sign  & Explanation & Value \\ \hline
$b$ & Minimal birth rate & $1.5\ast$ \\
$f$ & Increment of birth rate by wealth & $0.3\ast\ast$\\
$d$ & Decay rate of wealth & $0.5\ast\ast$\\
$N_f$ & Initial number of families in a society & $30\ast$\\
$N_s$ & Number of societies in a system & $50\ast$\\
$\mu $ & Mutation rate for $s$ and $\lambda$ & $0.03\ast\ast$\\
$\sigma$ & Strength of the noise in productivity & $0.1$\\ 
$c$ & Ratio of capacity and $N_f$ & Varied \\
$\epsilon$ & Wealth required for survival & Varied \\
$s$ & Probability to form extended family & Evolve \\
$\lambda$ & Inequality in the inheritance & Evolve
    \end{tabular}}
\end{table}

The life cycle of the families depends on the strategy $s$.
With the probability of $1 - s$, siblings are separated before agricultural production, i.e. they inherit some property determined by $\lambda$, lose $\epsilon$ of wealth, and build families of their own to produce independently. In contrast, with a probability of $s$, siblings remain in their parents' family to produce together, after which they are separated. According to the law of diminishing returns, productivity increases sub-linearly with labour force input \citep{ricardo1891principles, malthus1798essay, bacci2017concise}.
Following a study on pre-industrial farming \citep{evenson2001effect}, we assume that production increases in proportion to the logarithm of labour input, and is perturbed by Gaussian noise with a mean of $0$ and a variance of $\sigma^2$, resulting from internal and environmental fluctuations. 

Then, as long as the law of diminishing returns is satisfied, the total output of siblings is always higher when each sibling produces independently to form a nuclear family than when they concentrate their labour in an extended family. In this model, the productivities of $N$ members are $N \log 2$ and $\log (N + 1)$ for nuclear and extended families, respectively. However, under a limited capacity of available land resources, the total output of the society consisting of nuclear families will be lower than that of extended families, as a result of inefficient land usage (the productivities are $ \frac{1}{N} N \log 2$ and $\log (N + 1) $ for nuclear and extended families, respectively). In other words, there is a conflict between family- and society-level preferences for nuclear versus extended families under the conditions of limited capacity. As long as one considers pre-industrial farming, it is expected that family members work together on their land, and the law of diminishing returns is satisfied. However, this formulation will be inappropriate for modern farming or other subsistence patterns, which is beyond the scope of this model. The results for different formulations of labour-extensive subsistence patterns are briefly discussed.

After sibling separation and the production of wealth, each family reproduces. 
The number of children in families is positively correlated with their wealth in pre-industrial societies \citep{gibson2011land}. Here, we assume that it follows the Poisson distribution with a mean of $b + fw$, where $b$ and $f$ represent the minimal birth rate and the increment of birth rate by wealth $w$, respectively.
Children culturally inherit $s$ and $\lambda$ from their parents, with a slight variation through `mutation' at the rate of $\mu$, according to previous studies on cultural evolution \citep{cavalli1981cultural, creanza2017cultural}. At the time of altering generations, families lose $dw$ of wealth, where $d$ represents the decay rate of wealth due to ageing equipment, disaster, or taxation, for example.
Additionally, each society splits if the number of families reaches $2N_f$ at that time. The parameters are summarised in Table \ref{table:param}.

The above is a minimal model to discuss the diversification of family systems concerning parent--child and inter-sibling relationships. To consider husband--wife relationships, we extend the model by assigning each family male and female populations, and the strategy for inheritance distribution for sons and daughters. In the extended model, one can have multiple spouses by paying sufficient bridewealth. Polygyny increases both production and reproduction. This model is explained in the Supplementary Text in detail.

\subsection{Algorithm of the Model}
In this subsection, we show the mathematical formulation of our model.
We adopted the following algorithm for changes in the wealth and population of families. 
For the parent family $i$ and its $j$th child's family $i,j$, the population $N$ and the amount of wealth $w$ at time $t$ are expressed as follows:

 \begin{align}
 w_{i}^{t\ast} &= (1 - d) w_{i}^{t-1}. \label{eq:decay}
 \shortintertext{With probability $1 - s_i^t$,}
 N_{i,j}^t &= 1 \ \ (1\le j \le N_{i}^{t}), \label{eq:sep1}\\
 w_{i,j}^{t\ast} &= w_{i}^{t\ast} e^{-\lambda_i^t j} / \sum_{k = 1}^{N_i^t} e^{-\lambda_i^t k} - \epsilon, \label{eq:sep2}\\
 w_{i,j}^t &= w_{i,j}^{t\ast} + r(1 + \eta)(1 + w_{i,j}^{t\ast})\log(1 + N_{i,j}^t); \label{eq:prod1}
 \shortintertext{otherwise,}
 w_{i}^t &= w_{i}^{t\ast} + r(1 + \eta)(1 + w_{i}^{t\ast})\log(1 + N_{i}^t), \label{eq:prod2} \\
 w_{i,j}^{t} &= w_i^{t} e^{-\lambda_i^t j} / \sum_{k = 1}^{N_i^t} e^{-\lambda_i^t k} - \epsilon \ \ (1\le j \le N_{i}^{t}). \label{eq:sep3}
 \shortintertext{Then,}
N_{i,j}^{t+1} &= \text{Poisson}(b + f w_{i,j}^t), \label{eq:birth} \\
 s_{i,j}^{t+1} &= s_i^t + \zeta,\ \  \lambda_{i,j}^t = \lambda_i^t+ \zeta, \label{eq:mutation}
 \shortintertext{where} 
 r &= \min(1, cN_f / \text{\# families}), \label{eq:cap}\\
 \eta &\sim N(0, \sigma^2),\ \ \zeta \sim N(0, \mu^2). \label{eq:noise}
 \end{align}
 
In each simulation step (generation), families lose a proportion $d$ of their wealth (equation \eqref{eq:decay}). With probability $1-s$, siblings leave their parents' home (equation \eqref{eq:sep1}), distribute the inheritance with a wealth loss of $\epsilon$ (equation \eqref{eq:sep2}), and produce independently (equation \eqref{eq:prod1}).
Otherwise (with probability $s$), siblings produce together at their parents' home (equation \eqref{eq:sep3}) and build their own families after production (equation \eqref{eq:prod2}) instead of equations (\ref{eq:sep1}, \ref{eq:sep2}, \ref{eq:prod1}). Here, the production of wealth is inversely proportional to the number of families in society if the capacity is exceeded (equation \eqref{eq:cap}), is proportional to the logarithm of labour and increases with linear feedback from wealth, and is perturbed by noise resulting from internal and environmental fluctuations following a normal distribution (equation \eqref{eq:noise}). 
Finally, families produce offspring (equation \eqref{eq:birth}) and strategy parameters are transmitted with slight mutation (equation \eqref{eq:mutation}). The birth rate increases linearly with wealth. Here, families reproduce without considering marriage explicitly. The extended model with the marriage process is explained in the Supplementary Text.

In the simulation, the initial values of strategies are $s = 0.5$ and $\lambda = \log 2$ for all families. 
Hence, at the initial state, families can form nuclear or extended families with equal probability, and the inheritance is moderately biased so that the share of the inheritance received by subsequent children is half of that received by the preceding children.
In other words, the families are not differentiated as nuclear or extended, or as equal or strongly biased inheritance providers.
However, no qualitative changes are observed under other initial conditions. The source code has been made publicly available in the Dataverse repository \citep{itao_data}, \url{https://doi.org/10.7910/DVN/3ZGCQI}.

\section{Results}
\subsection{Evolution of Family Systems}
 \begin{figure*}[tb]
    \centering
    \includegraphics[width=0.8\linewidth]{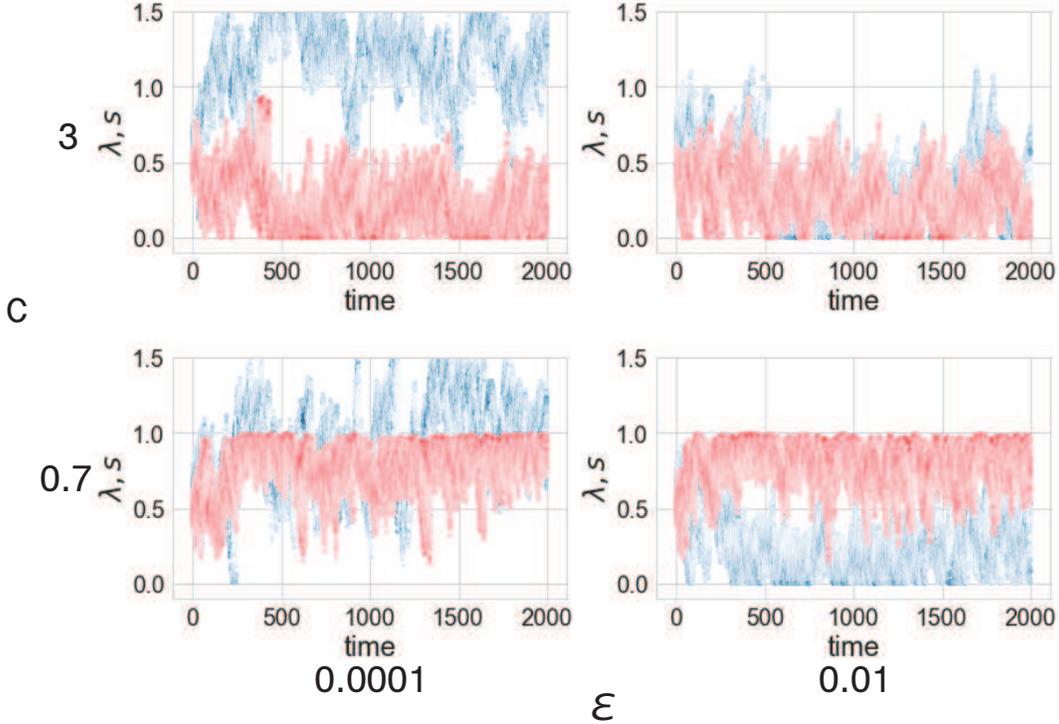}
    \caption{Time series of family strategy parameters. The temporal evolution of the values of the probability to form extended families $s$ (red) and the inequality in inheritance $\lambda$ (blue) of families in a society are plotted, for environmental conditions with $c = 0.7, 3$ and $\epsilon = 0.0001, 0.01$.}
    \label{fig:strat_dist}
\end{figure*}

\begin{figure*}[tb]
   \centering
   \includegraphics[width=\linewidth]{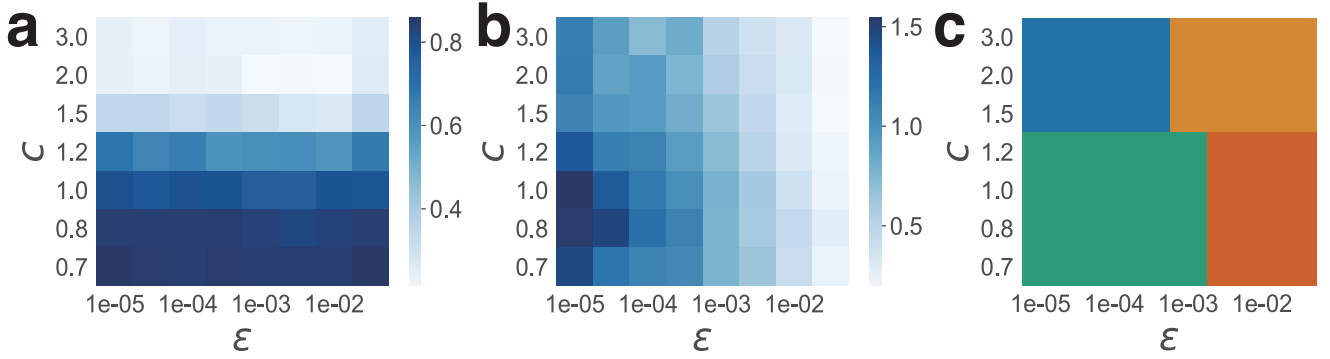}
    \caption{Dependence of family strategy parameters on environmental parameter values. The graphs show the average values of family strategies (\textbf{a,} the probability to form extended families $s$ and \textbf{b,} the inequality in inheritance $\lambda$) in the last 1,000 steps of simulation. 
    \textbf{c,} Phase diagram of family systems. The environmental dependencies of stem (green), community (orange), absolute nuclear (blue), and egalitarian nuclear (yellow) families are shown. Parent--child relationship for residence is classified as that of an extended family if $s \ge 0.5$ and as that of a nuclear family if $s < 0.5$. Inter-sibling relationship for inheritance is classified as unequal if $\lambda \ge \log 2$ and as equal if $\lambda < \log 2$.}
    \label{fig:phase_family}
\end{figure*}

\begin{figure}[tb]
    \centering
     \includegraphics[width=\linewidth]{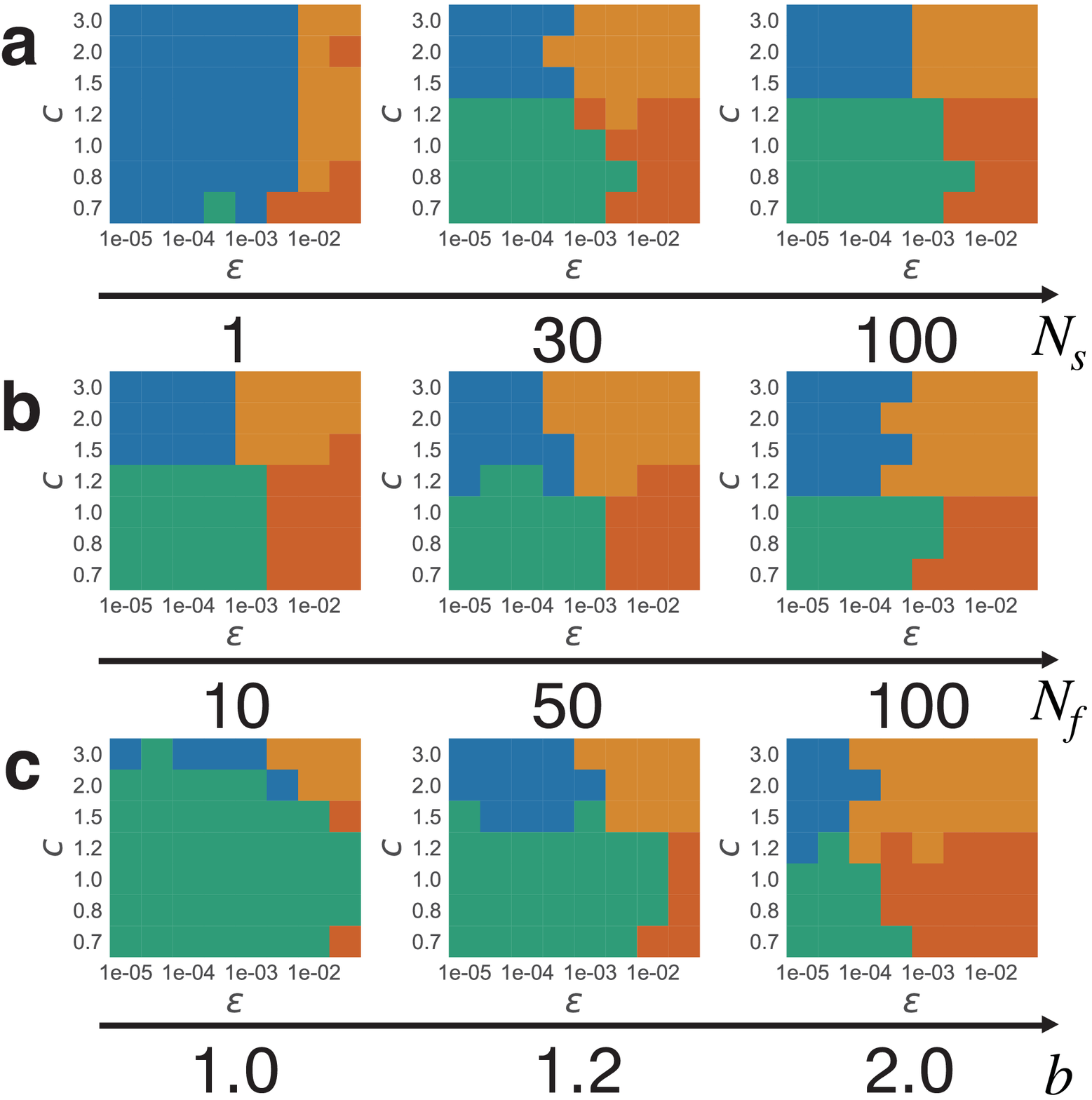}
    \caption{Dependence of the phase diagrams of family systems on parameters with $\ast$ in Table \ref{table:param} (\textbf{a,} the number of societies $N_s$ \textbf{b,} the number of families $N_f$ \textbf{c,} birth rate $b$). Unless the value is shown on the axis, the parameter values are fixed to those in Table \ref{table:param}.}
    \label{fig:phase_phase_structure}
\end{figure}

The simulations are performed for 2,000 time steps.
The time series of the evolution of family strategy parameters are shown in Figure \ref{fig:strat_dist}. Family strategies do not diverge within each society, but are concentrated around a specific value adapted to each given environmental condition.
The probability to form extended families $s$ (plotted in red) and the inequality in inheritance $\lambda$ (plotted in blue) evolve depending on the values of land capacity $c$ and wealth required for survival $\epsilon$, respectively. The evolution of the strategy converges within approximately 1,000 steps in every parameter region.

We conducted multi-level evolutionary simulation 100 times for each condition and averaged the strategy parameters of families in the last 1,000 steps. Figure \ref{fig:phase_family}a,b show the dependence of $s$ and $\lambda$ on $c$ and $\epsilon$. 
Increasing land capacity $c$ makes nuclear family strategies more preferable at both the family- and society-levels.
Then, the probability of forming an extended family $s$ decreases, implying the evolution of nuclear families.
Increasing $\epsilon$, the wealth required for survival, increases the demand for inheritance by younger siblings.
This results in a smaller inequality in inheritance $\lambda$, indicating the evolution of the equal inheritance.
Figure \ref{fig:phase_family}c shows the phase diagram of family systems. Here, we classify the family systems as extended if $s \ge 0.5$, and as nuclear if $s < 0.5$. Similarly, we classify them as unequal inheritance sharers if $\lambda \ge \log 2$ and as equal if $\lambda < \log 2$. This criterion is determined according to whether the share of the inheritance awarded to the subsequent children is smaller (or larger) than half of that awarded to the preceding children. Therefore, the four basic family systems evolve depending on the two environmental parameters $c$ and $\epsilon$. A stem family (plotted in green) evolves if $c$ and $\epsilon$ are small. A community family (plotted in orange) evolves if $c$ is small and $\epsilon$ is large. An absolute nuclear family (plotted in blue) evolves if $c$ is large and $\epsilon$ is small, and an egalitarian nuclear family (plotted in yellow) evolves if both $c$ and $\epsilon$ are large.

We show the dependence of the phase diagrams of family systems upon other parameters in Figure \ref{fig:phase_phase_structure} and Figure S1. The phase diagrams plotted against $c$ and $\epsilon$ are qualitatively robust and independent of the other parameter values. However, quantitative trends exist. Generally, $N_s$ and $N_f$ determine the strength of selection pressures at the family- and society-levels \citep{traulsen2006evolution, takeuchi2017origin}.
For a large $N_f$ or small $N_s$ (ultimately, if $N_s = 1$), family-level competition becomes dominant rather than society-level competition, which leads to the evolution of selfish behaviour. Conversely, when society-level competition is dominant because of small $N_f$ or large $N_s$, cooperative behaviour evolves.
Figure \ref{fig:phase_phase_structure}a,b show that, if $N_f$ is large or $N_s$ is small, nuclear families evolve even when $c$ is small. Recall that the total production of siblings increases if they work independently, but that of society decreases because of the inefficiency in land usage if the capacity is limited. Therefore, choosing a nuclear family under small $c$ is a selfish strategy that evolves for small $N_s$ and large $N_f$. 
As wealth $w$ accumulates, $\epsilon$ would become relatively small for the wealth, and accumulation would be accelerated.
However, when the minimal birth rate $b$ is high, the number of offspring increases, and each share decreases. Therefore, less wealth is accumulated, and equal distribution evolves in larger parameter regions because of the relatively large $\epsilon$, as shown in Figure \ref{fig:phase_phase_structure}c. The dependence of the phase diagram on the parameters mutation rate $\mu$, decay of wealth $d$, and increment of birth rate by wealth $f$ are shown in Figure S1.

\subsection{Evolution of Husband--Wife Relationships in the Extended Model}
The simulation results are shown in Figure S2 for the extended model considering the marriage process. Parental investment is biased for sons almost four times as much as daughters for most of the parameter regions. Investment for sons is advantageous because wealthy men can have many wives and increase their fitness.
Evolutionary anthropologists have reported that daughter-biased investment evolves under paternity uncertainty \citep{holden2003matriliny}, which is beyond the scope of our model. Hence, it is reasonable that only son-biased investment evolves in our model.

Results also show that the frequency of polygyny is less than 20 \%. The bias of parental investment and the frequency of polygyny are almost independent of land capacity $c$ or wealth required for survival $\epsilon$.
Furthermore, consideration of the marriage process only minimally affects the results for parent--child and inter-sibling relationships. Hence, we will analyse the economic structures of evolved societies by using the previous minimal model in the following section.

Additionally, we studied a model in which the diminishing returns of family labour in production are relaxed, to consider labour-extensive subsistence patterns other than agriculture.
Figure S3 shows that if production increases linearly to the number of family labourers, the frequency of polygyny increases to more than 20 \% almost independently of $c$ and $\epsilon$, even though the parental investment bias is almost the same as the above model.
In this model, the increase in polygyny results from a larger fraction of relatively wealthy people. This scenario is consistent with empirical reports for foraging, horticultural and agropastoral societies \citep{ross2018greater}.
Extended families are dominant even when land capacity $c$ is sufficient because they are no less preferable than nuclear families even at the family-level in this model. However, both nuclear and extended families are observed in most of the subsistence patterns \citep{murdock1969standard}.
It suggests that nuclear families can evolve owing to some reasons not covered by our model. To discuss the variation of family systems depending on subsistence patterns, it will be necessary to consider the difference in productivity and lifestyle.

\subsection{Wealth Distribution and Evolution of Social Structure}
\begin{figure*}[tb]
    \centering
    \includegraphics[width=1.0\linewidth]{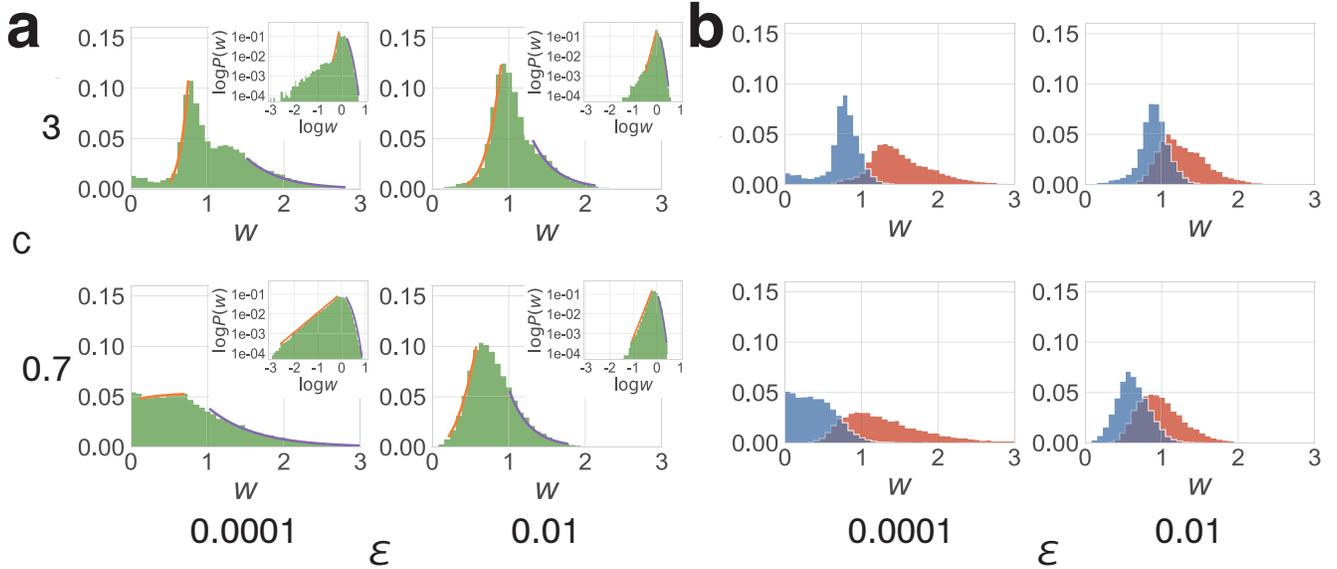}
    \caption{Distributions of wealth in evolved societies. \textbf{a,} The graphs show the frequency distribution of family wealth in the last 1,000 steps for environmental conditions with $c = 0.7, 3$ and $\epsilon = 0.0001, 0.01$. The orange line shows the power-law fitting, and the purple line shows the exponential distribution fitting. The insets show the log--log plot of the frequency distribution. \textbf{b,} The wealth distribution of the eldest (red) and younger siblings (blue).}
    \label{fig:wealth_dist}
\end{figure*}

\begin{figure*}[tb]
    \centering
    \includegraphics[width=0.75\linewidth]{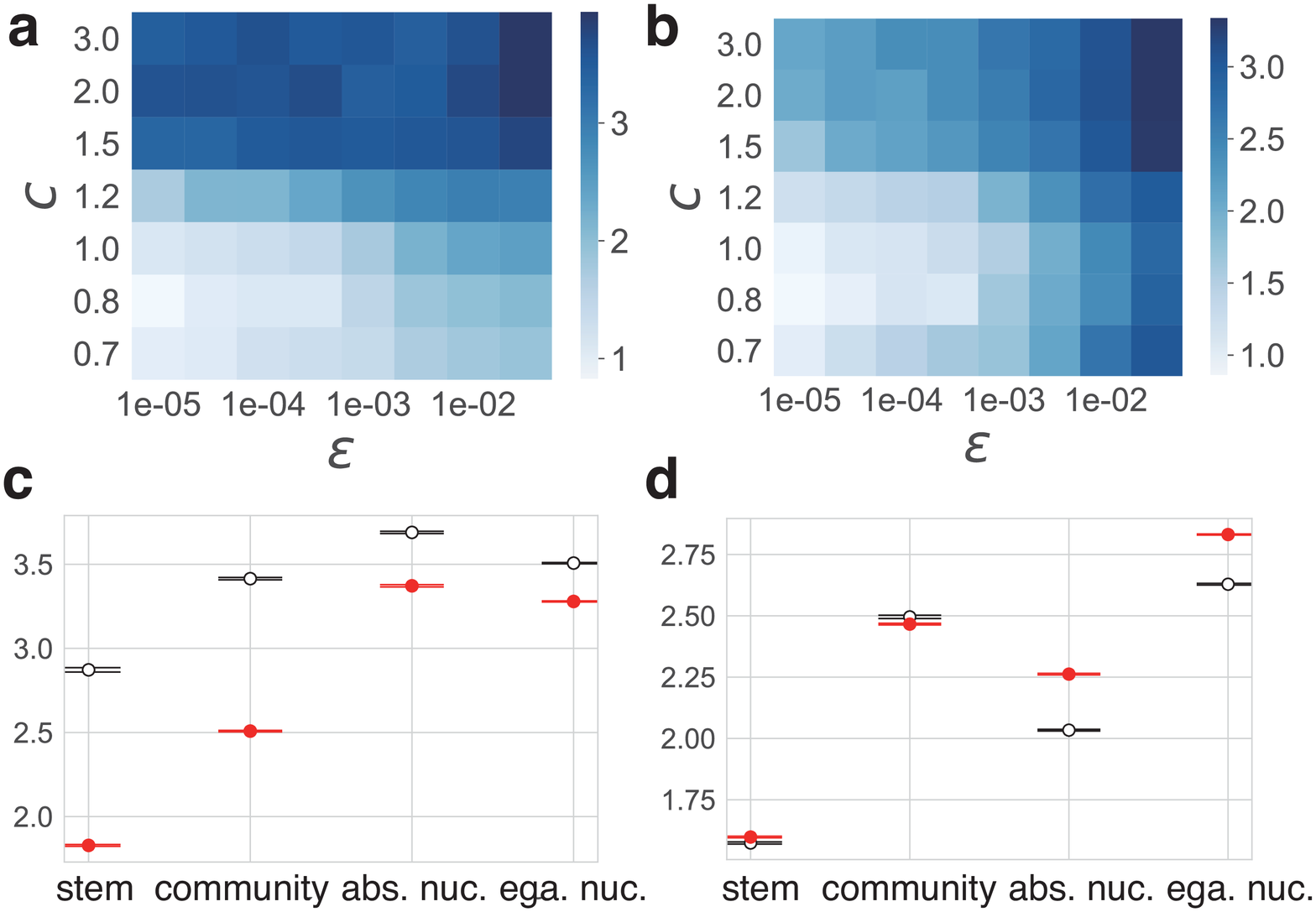}
    \caption{Dependence of the heaviness of tails of wealth distribution on environmental parameters and family systems. \textbf{a,} Lightness of the poor tail $\alpha$ given by the power-law fitting $w^\alpha$. \textbf{b,} Lightness of the rich tail $\beta$ given by the exponential fitting $\exp(-\beta w)$. Smaller $\alpha$ and $\beta$ show heavier tails for the poor and rich sides, respectively.
    \textbf{c, d,} Averaged value of $\alpha$ (\textbf{c}) and $\beta$ (\textbf{d}) classified by the dominant family system (stem, community, absolute nuclear, and egalitarian nuclear families) in each society, with error bars. Values for the environmental parameters fixed to $\epsilon = 0.0003$ and $c = 1.5$ are plotted in black, whereas those averaged over $\epsilon = 0.0001, 0.0003, 0.001, 0.003$ and $c = 1.0, 1.2, 1.5, 2.0$ are plotted in red. 
    }
    \label{fig:phase_index}
\end{figure*}

Subsequently, we investigated the wealth distribution of families for each society after evolution.
Note that data from the wealth distribution in modern society suggest an exponential-type tail for the rich side \citep{chakrabarti2013econophysics, tao2019exponential} (say a log-normal \citep{gibrat1931inegalits} or gamma distribution \citep{chakraborti2008gamma}), and a power distribution for the poor side \citep{reed2003pareto} (say a gamma distribution \citep{chakraborti2008gamma}). The gamma distribution is obtained by assuming the wealth growth with positive feedback and nonlinear saturation, as well as a multiplicative stochastic process, which are included in our model (see Supplementary Text).

Figure \ref{fig:wealth_dist}a shows the frequency distribution of the wealth of families within each society. In every parameter region, the distribution approximately follows a power-law on the poor side and has an exponential tail on the rich side, which is consistent with the above data.
Because inherited wealth depends on birth order, we also plotted the distributions of wealth by distinguishing the eldest siblings from the others in Figure \ref{fig:wealth_dist}b.
With decreasing wealth required for survival $\epsilon$, and consequently increasing inheritance inequality, the distributions of the wealth of siblings separate further. As a result, the accumulation of wealth by heirs is accelerated. Decreasing land capacity $c$ and the evolution of extended families result in poorer younger siblings, whereas greater land capacity $c$ and the evolution of nuclear families give rise to wealthier younger siblings.

Although the wealth distribution follows the power-law $w^\alpha$ on the poor side and the exponential distribution $\exp(-\beta w)$ on the rich side universally, the heaviness of tails depends on the environmental parameters. We fitted values for the lightness of the poor tail $\alpha$ and those for the rich tail $\beta$ and averaged them over 100 trials for each environmental parameter $c$ and $\epsilon$ in Figure \ref{fig:phase_index}a,b, respectively.
Smaller $c$ results in smaller $\alpha$, i.e. the heavier tail is on the poor side, while smaller $\epsilon$ results in smaller $\beta$, i.e. the heavier tail is on the rich side.
These results suggest the characteristics of the wealth distribution in the four corresponding family systems. However, it remains unclear whether they result from environmental conditions or family systems. To confirm the relevance of family systems, we computed the dependence of wealth distributions on family systems by sampling each family system using fixed environmental parameters near the boundary of the four phases of family systems, where the values of $s$ and $\lambda$ are distributed to cover all four family systems.

The average values of $\alpha$ and $\beta$ for each family system, which were sampled for the fixed environmental parameters (plotted in black), are shown in Figure \ref{fig:phase_index}c,d. 
They demonstrate the trend that the poor tail is heavier for extended families with poor younger siblings, whereas the rich tail is heavier for unequal inheritance with rich heirs. By comparing these results with the results averaged over several environmental parameters around the phase boundary (plotted in red), it is shown that the above trend depends on each family system, and is further intensified by environmental parameter values.

\subsection{Empirical Data Analyses}
Next, we verify our results on the relationship between environmental conditions, family systems, and economic structures. Using the global ethnographic database of 186 premodern societies, called SCCS \citep{murdock1969standard, kirby2016d}, empirical data analyses were conducted.

First, we classified family systems of pre-industrial agricultural societies. We then identified pre-industrial agricultural societies by using the \textit{Subsistence economy: dominant activity} variables (5, 6, and 7 correspond to agriculture).
Then, we identified the family systems by using \textit{Domestic organization} (1, 2, 3, 4, and 5 correspond to nuclear families and 6, 7, and, 8 correspond to extended families) and \textit{Inheritance distribution for movable property} (1 corresponds to equal and 2, 3, and 4 correspond to strongly biased inheritance). (See Supplementary Tables for a detailed explanation of these variables.)
Out of 186 societies in SCCS, 91 societies conducted agriculture and inheritance of movable properties.
Among them, 14 societies were classified as stem families, 30 as community families, 17 as absolute nuclear families, and 30 as egalitarian nuclear families. Figure S4 shows their geographic distribution.
Here, we used the data on the inheritance of movable properties to identify inter-sibling relationships. However, similar trends on the following variables were achieved, even when we used those pertaining to the inheritance of real property, as shown in Table S2.

Next, we conducted Spearman's rank correlation analyses and calculated the correlation between SCCS variables and parent--child (nuclear or extended) or inter-sibling (strongly biased or equal) relationships. 
The database contains various variables of socio-ecological factors. By calculating the correlation for each variable and listing the variables in descending order in the absolute value of the correlation, we found those variables related to the parameters in our model in the top of the list.
The variables that are highly correlated with parent--child and inter-sibling relationships are listed in Table S1 and Table S2, respectively. Among them, we show the variables that can be related to the model parameters in Table \ref{table:corr}. 

Table \ref{table:corr} shows the dependence of family systems on environmental conditions. \textit{Communality of land} and \textit{Land Shortage}, suggesting larger and smaller land capacity $c$, respectively, are correlated with extended families (Corr. -0.31 (P = 0.03) and Corr. 0.26 (P= 0.09), respectively). This is consistent with the theoretical results showing the evolution of extended families for smaller $c$.
On the other hand, \textit{Frequency of internal warfare} and \textit{Acceptability of violence within society} suggest that violent conflict is more frequently observed in societies with equal inheritance (Corr. 0.36 (P = 0.08) and Corr. 0.27 (P= 0.13), respectively).
Such violence will damage goods and require families to have more wealth to survive; as a result, the wealth required for survival $\epsilon$ increases in our model. Accordingly, equal inheritance is more frequent for larger $\epsilon$, as is consistent with our results.

Furthermore, the data suggest the correlation between family systems and economic structures. \textit{Number of poor} implying smaller $\alpha$ is positively correlated with extended families (Corr. 0.29 (P = 0.06)), whereas \textit{Number of rich people} implying smaller $\beta$ is negatively correlated with equal inheritance (Corr. -0.37 (P = 0.01)). Thus, the empirical data are consistent with our simulation results, concerning the relationship between environmental conditions, family systems, and society-level economic structures.
See Supplementary Tables for the explanation on values of SCCS variables.

\begin{table}[tb]
\caption{Correlation of variables in SCCS and parent--child (nuclear or extended) or inter-sibling (unequal or equal) relationships. The correlated variables, their correlation, and the corresponding parameters in our model are presented. Here, we list the variables relevant to the model. The three variables above the dashed line are correlated with the parent--child relationship, and those below the dashed line are correlated with the inter-sibling relationship. See Supplementary Tables for the list and detailed explanation of the variables with high correlation. *P $< 0.1$, **P $< 0.05$, ***P $< 0.01$.} 
  \label{table:corr}
 \centering
    \begin{tabular}{lll} 
Variable	&	Corr.	&	Model	\\\hline
Communality of land	&	-0.31**	&	$c$	\\
Number of poor	&	\ 0.29*	&	$\alpha$	\\
Land shortage	&	\ 0.26*	&	$c$	\\\hdashline
Number of rich people	&	-0.37***	&	$\beta$ \\
Frequency of internal warfare	&	\ 0.36*	&	$\epsilon$	\\
Acceptability of violence	&	\ 0.27	&	$\epsilon$
    \end{tabular}
\end{table}

\section{Discussion}
By simulating the multi-level evolution model of family systems, we demonstrated the evolution of family systems depending on the environmental parameters for the capacity of available land resources $c$ and the amount of wealth required for a family to survive $\epsilon$. As for parent--child relationships, if there is sufficient land capacity, nuclear families evolve, whereas extended families evolve under a land shortage. As for inter-sibling relationships, if the wealth required for survival is large, equal inheritance evolves, whereas strongly biased inheritance evolves when that is small.
Therefore, the four basic family systems characterised by both relationships above are represented as `phases' depending on $c$ and $\epsilon$.
By considering marriage, we then confirmed son-biased investment and infrequent polygyny.
Additionally, we clarified the characteristics of wealth distribution determined by the dominant family systems within societies. The tail of the poor side is heavier (that is, many poor people) for extended families, and that of the rich side is heavier (that is, many rich people) for families with unequal inheritance. Empirical data analyses of premodern societies in SCCS supported our results. 

Now, we refer to demographics and socioeconomic histories in the premodern and modernising era, especially in Western Europe and East Asia.
The land capacity $c$ in our model can be measured approximately by the period since the onset of agriculture. In the areas where agriculture started early, population growth resulted in the exhaustion of available land, and labour-intensive farming developed, as observed in China \citep{wallerstein2011modern, pomeranz2000great}, Russia \citep{hizen1994demographic}, and Japan \citep{hayami2015japan}. 
In Western Europe, especially Holland, the Paris Basin, Southern England, and Central Spain, the capacity was large until industrialisation, and labour-saving farming was developed \citep{pomeranz2000great} as a result of the following reasons: agricultural progress in medieval times enabled virgin land cultivation by gathering the children not inheriting the parental lands \citep{grigg1980population, pirenne1956economic, cameron1993concise, bacci2017concise}; the population stagnated in premodern times because of religious wars and plagues \citep{bacci2017concise}; and colonies were established early on \citep{wallerstein2011modern}. 
Accordingly, the model result concerning $c$ implies that extended families evolve in the areas where agriculture started early.
Table \ref{table:corr} also suggests that the exhaustion of land leads to the evolution of extended families. 

As for wealth required for survival $\epsilon$, demographics report that the frequency of violent conflict decreased in the following order in Eurasia: the centre of the continent, peripheral, and island regions \citep{macfarlane2002savage, khazanov2012nomads, umesao2003ecological}. The regions close to the pole of civilisation and/or those frequently attacked by foreign people would have a large $\epsilon$. 
Hence, the model result concerning $\epsilon$ implies that equal inheritance is dominant in the centre of the Eurasia continent and other regions that are vulnerable to warfare (see Figure S4).
The results of the empirical data analysis in Table \ref{table:corr} also support the correlation between such violent conflict with the evolution of equal inheritance. 

From geohistorical reports discussed above, the family systems in each region can be explained: absolute nuclear families (nuclear family, unequal inheritance) in England and the Netherlands, where available land resources were sufficient and wealth required for survival was small; egalitarian nuclear families (nuclear family, equal inheritance) in France, Spain, and Italy, where both land capacity and necessity of wealth were large; stem families (extended family, unequal inheritance) in Japan, Germany and many parts of rural Western Europe, where both of them were small; and community families (extended family, equal inheritance) in China, Russia, and Northern India, where land capacity was small and necessity of wealth was large \citep{berkner1972stem, todd2011origine}.

Apart from these environmental conditions, the number of families within a society $N_f$, the number of competing societies $N_s$, and birth rate $b$ are also relevant parameters for determining the family system. $N_f$ is large for large-scale land management as seen in England, the Netherlands, France, and Spain, whereas $N_f$ is small and $N_s$ is large in family farm management as observed in China, Russia, Japan, and Germany \citep{pomeranz2000great, wallerstein2011modern, hizen1994demographic, cameron1993concise, bacci2017concise}.
The trends in Figure \ref{fig:phase_phase_structure}a,b are consistent with the observation of nuclear families in the former regions and extended families in the latter.
The birth rates were low in Japan and Western Europe, especially in England \citep{macfarlane2002savage}, and higher in Russia \citep{hizen1994demographic}. The observation of unequal inheritance in the former regions and equal inheritance in the latter demonstrates a similar tendency to Figure \ref{fig:phase_phase_structure}c.

Next, we examine the validity of our results regarding the relationships between family systems, socio-economic structures, and modern social ideologies.
Figure \ref{fig:phase_index} suggests that, in England and the Netherlands involving absolute nuclear families, the tail of wealth distribution is heavy on the rich side and light on the poor side. Indeed, wealthy farmers prospered and employed a majority as labourers who had better living standards than those in poorer regions \citep{wallerstein2011modern, todd1990invention, tawney1912agrarian, shaw2012rise, macfarlane2002savage, laslett2015world}.
The accumulation of capital and independent labour forces explains the development of individual liberty and capitalism in England \citep{wallerstein2011modern, braudel1992civilization1, braudel1992civilization2, braudel1992civilization3, todd1990invention}.
Wealth distribution in France and Spain, involving egalitarian nuclear families, was suggested to have light tails on both the rich and poor sides. That is, agricultural societies were less differentiated and weakly stratified \citep{wallerstein2011modern, rosener1993bauern, dupeux1972societe}, which forms the basis of the values of freedom and equality.
Our results suggest that wealth distribution in Germany, Sweden, and Japan, involving stem families, had heavy tails on both rich and poor sides.
Wealthy farmers prospered by exploiting others and the stratification of society advanced in accordance with the order and class distinctions \citep{todd1990invention, rosener1993bauern, kastner1978chronik, mager1981haushalt, hayami2001regional, hayami2015japan}.
Wealth distribution in Russia and China, involving community families, was suggested to have a light rich tail and a heavy poor tail. Indeed, the middle class was significantly sparse, and people were uniformly poor \citep{rosener1993bauern}, which led to the adoption of communism \citep{weber1995russian, thaxton1997salt}. In this manner, the wealth distribution obtained in our model connects family systems with society-level characteristics observed in socio-economic history. 
A study of political ideology showed that people supported authoritarianism in the presence of many individuals being exposed to threats, and egalitarianism in the absence of strong inequality or power imbalance \citep{claessens2020dual}.
Our results are consistent with this, because the heavier poor and rich tails imply the presence of vulnerable and privileged people, respectively.

Note that our results regarding the family systems and the socio-economic structures are expected to be rather general. The conclusion here is independent of the details of the present model, as long as the production increases sub-linearly with labour input, and multi-level selection of families and societies is considered.

One can also discuss long-term changes in land capacity $c$ and wealth required for survival $\epsilon$, and their influence on family systems.
At times of cultivation, a nuclear family evolves because of sufficient capacity. As the population increases and capacity becomes limited, an extended family would replace it.
Additionally, because of the dense population, the risks of invasion from surrounding areas and conflict within societies increase, and accordingly, owing to the loss of wealth by violence, $\epsilon$ would increase gradually.
This scenario explains the historical change of family systems from a nuclear family to a stem family, and then to a community family \citep{todd2011origine}.

Environmental factors change gradually owing to the interaction between society and the environment. However, the change in environmental factors, in turn, will alter family systems and social structures. Such historical dynamics have been discussed as the interaction of factors on different time scales \citep{braudel1992civilization1, braudel1992civilization2, braudel1992civilization3}. To discuss such interaction of factors at different levels, the present constructive model will give a basic explanation.

The present model has some limitations. First, the differentiation of people between the elite and the majority was not considered. As society becomes stratified, people start to rent land from the elite. This results in the divergence of environmental factors and family systems between them \citep{todd2011origine}. A model is needed for handling social stratification and the interaction of classes to discuss broader issues.
Second, we did not model intra-family competition for resources explicitly and the relationships between family systems and such competition remain unsolved. Intra-family competition has attracted attention in evolutionary anthropology \citep{trivers1974parent, ji2014intergenerational}. In fact, the previous studies have revealed the high status of the elderly in extended families \citep{lee1979family}, and sibling competition over the land inheritance \citep{gibson2011land}.
The model, then, needs to have three levels, i.e. individual, family, and society.
Finally, the current model focuses only on pre-industrial agricultural societies. Models focusing on other subsistence patterns are needed to discuss the diversity of family systems widely. For example, Figure S3 shows that if the diminishing returns of family labour are relaxed, the frequency of polygyny increases.
Furthermore, in the modern world, agricultural societies should no longer be regarded as isolated systems, but constitute components of a world-system \citep{wallerstein2011modern}. A new model needs to be developed to consider the interaction between towns and agricultural societies, as well as international, political, and commercial networks.

Furthermore, the present empirical data analyses have some limitations. 
First, we could only analyse the correlation between cultural variables and family systems. Although it is desirable to conduct better analyses (such as classification learning) to reveal features relevant to family systems, it was infeasible due to data insufficiency. 
In this study, we used SCCS in which the data for a variety of socio-ecological factors are available.
SCCS enabled us to test the correlation between model parameters and family systems empirically, although the sample size was limited.
Future works should make use of other databases that include more societies (but fewer variables per society), such as the Ethnographic Atlas \citep{murdock1967ethnographic}.
Second, we could not analyse causal relationships owing to a lack of chronological data. Our correlation analyses are insufficient to examine whether history progresses as our model predicted.

Here, the collaboration of field studies, historical analyses, and theoretical modelling is necessary to further elucidate the historical dynamics of societies.
Field studies describe the individual- or family-level behaviour and society-level structures synchronically. Historical or phylogenetic analyses unveil the diachronic change of such factors. 
In addition, Leach has emphasised the importance of generalising ethnographic findings by using mathematical formulation to unveil universal structural patterns that may appear in any type of society \citep{leach1961rethinking}.
The constructive model, as presented here, provides a simplified mathematical expression of family behaviours, gives a general framework that allows comparison of various societies, explains the universal patterns between family- and society-level factors, and reveals their historical evolution.

To summarise, we presented a multi-level evolution model to account for the emergence of the four basic family systems and the resultant socio-economic structures depending on environmental conditions, as is consistent with family-level anthropological studies and society-level economic histories. Here, the microscopic characteristics of families determine the macroscopic economic structures, which forms the basis for the development of societies. This study allows an explanation of the universal evolutionary constraint that human societies satisfy.

\subsubsection*{Data availability}
 The source code has been made publicly available in the Dataverse repository \citep{itao_data}, \url{https://doi.org/10.7910/DVN/3ZGCQI}, or \url{https://github.com/KenjiItao/family\_system.git}.

\subsubsection*{Competing interests}
The authors declare no conflict of interest.

\subsubsection*{Funding}
This study was partially supported by a Grant-in-Aid for Scientific Research on Innovative Areas (17H06386) from the Ministry of Education, Culture, Sports, Science, and Technology (MEXT) of Japan.

\subsubsection*{Acknowledgements}
The authors thank Tetsuhiro S. Hatakeyama, Yuma Fujimoto, and Kenji Okubo for a stimulating discussion, and Yasuo Ihara for illuminating comments.

\end{document}